\documentclass{article}

\usepackage[]{graphicx}

\newcommand{\be}{\begin{equation}}
\newcommand{\ee}{\end{equation}}

\begin{document}


\title{Hubble constant from lensing in plasma-redshift cosmology, and intrinsic redshift of quasars}         
\author{Ari Brynjolfsson \footnote{Corresponding author: aribrynjolfsson@comcast.net}}

\date{\centering{Applied Radiation Industries, 7 Bridle Path, Wayland, MA 01778, USA}}          

\maketitle

\begin{abstract}  In a series of articles, we have shown that the newly discovered plasma-redshift cosmology gives a simpler, more accurate and consistent explanation of many cosmological phenomena than the big-bang cosmology.  The SNe Ia observations are in better agreement with the magnitude-redshift relation predicted by the plasma redshift than that predicted by the multi-parameter big-bang cosmology.  No deceleration or expansion parameters are needed.  The plasma-redshift cosmology is flat and quasi-static on a large scale.  The Hubble constant is no longer an expansion parameter, but is instead a measure of the average electron density along the line of sight towards an object.  Perusal of the SNe Ia data and quasar data has shown that there is no time dilation.  The conventional estimates of the Hubble constant from gravitational lensing observations use the big-bang cosmology for interpreting the observations.  This has lead to a large spread and discordant estimates of the Hubble constant.  The purpose of the present article is to show that the gravitational lensing observations are in agreement with the plasma-redshift cosmology, and to show how to evaluate the lensing observations based on the new plasma-redshift cosmology.  The lensing observations also indicate that the quasars have large intrinsic redshifts.

\end{abstract}




\tableofcontents

\makeatletter	   
\renewcommand{\ps@plain}{
     \renewcommand{\@oddhead}{\textit{Ari Brynjolfsson: Hubble constant from lensing and the time dilation in a plasma universe}\hfil\textrm{\thepage}}%
     \renewcommand{\@evenhead}{\@oddhead}
     \renewcommand{\@oddfoot}{}
     \renewcommand{\@evenfoot}{\@oddfoot}}
\makeatother     

\pagestyle{plain}


\vspace{14mm}

\section{Introduction}

In references [1-4], we have shown that a newly discovered plasma redshift gives a simple and beautiful explanation of a long series of observations that have been difficult to explain.  The plasma redshift is caused by interaction of photons with hot and sparse plasma.  In the plasma redshift the photons transfer energy to the hot, sparse plasma in extremely small quanta, but the cross section for each individual interaction is very large.  The cross section for the plasma redshift has been overlooked because the conditions, low density and high temperature over extended space that are required for its detection, are difficult to create in the laboratory; but these conditions are ubiquitous in space.  In ordinary laboratory plasma, the cross section is zero.

\indent  The part of the photon energy that is lost in the plasma redshift, is transferred to the plasma and causes significant heating of the plasma.  In the Sun the plasma redshift heating starts low in the transition zone to the corona. The plasma redshift, together with the conversion of magnetic field to heat, is the main cause for the steep rise of the temperature in the transition zone.  Even without the magnetic field, the plasma redshift alone is able to create a steep temperature rise in the transition zone to million degrees K in the corona [1] (see section 5.1 and 5.2 of that source).  The conversion of magnetic field to heat is usually induced and enhanced by the plasma-redshift heating [1] (see Appendix B of that source).

\indent  The plasma redshift also explains the heating of the coronas of the stars, the quasars, and the galaxies, and it explains the heating of the hot intergalactic plasma [1] (see sections 5.5 to 5.9 of that source).  In the past, the intergalactic space was surmised cold and empty, because we had no means for heating the intergalactic plasma.  The coronas of galaxies were surmised to have very low densities, because we could not find enough supernovae (or any other means) for heating the coronas.

\indent  Besides the plasma-redshift heating, the plasma redshift explains the main fraction of the redshifts of the Fraunhofer lines in the Sun.  Importantly, it explains in a simple way the cosmological redshift in intergalactic space [1] (see in particular section 5.8 of that source).  The Doppler shifts caused by movements of the emitting layer can often be distinguished and evaluated separately.

\indent  The deduction of the plasma redshift is given in reference [1] (see in particular sections 2, 3, and 4, and Appendix A of that source).  The deduction of plasma redshift is based on basic and well-proven axioms of conventional physics without any new assumptions.  It is unfortunate that this plasma-redshift cross-section was not discovered 60 years ago, when it could have been discovered.  Similar calculations have lacked adequate exactness, and this atypical cross-section has been overlooked.

\indent  In reference [2], we show that the SN\,Ia data are consistent with {\it{no time dilation}}.  This contradicts the widely held belief that SNe\,Ia data show time dilation.  A SNe\,Ia is not a perfect standard candle.  The Malmquist bias causes a significant increase in absolute magnitude with distance.  The absolute brightness of the SNe\,Ia increases with the width of the light curve and with the distance and therefore with 1+z.  The part of the dimming that has been assumed to be caused by the time dilation is actually caused by one half of the Compton effect [2].  The other one half of the Compton effect masks the Malmquist bias.  This is all supported by robust data of excellent quality.

\indent  The SNe\,Ia data are consistent with the magnitude-redshift relation predicted by the plasma-redshift cosmology [1,\,2,\,3].  The distance to an object is given by $R = (c/H_0)\,{\rm{ln}}\,(1+z).~$  The observed light intensity decreases proportional to $1/R^2 .~$  The Compton scattering reduces the bolometric light intensity by a factor of $1/(1+z)^2 ,$ and the redshift of the photons reduces the intensity by $1/(1+z) .~$  The light intensity is thus proportional to  $1/[R^2 (1+z)^3] .~$  There is no time dilation, no deceleration, and no expansion of the universe.  The agreement between predictions of the plasma-redshift cosmology and the magnitude of SNe Ia is excellent.  No variable and adjustable parameters, such as $\Omega_m,~\Omega_{\Lambda},{\rm{and}}~\Omega_k ,$ are needed.

\indent  The observations indicate that the universe is quasi-static, everlasting, and infinite.  The universe is everlasting, because matter can through conventional processes renew itself [1] (see in particular section 6 of that source).  Although the universe is infinite, the gravitational potential is finite.  This is a quantum mechanical effect [1,\,4].  According to quantum mechanics, it takes a finite time for a weak gravitational field to transfer information about its direction and strength to a particle.  During this finite time the distant particle is being bombarded by the surrounding particles in the hot environment of the plasma universe.  The gravitational field therefore becomes inactive at very large distances.  Apart from small ripples (as in the gravitational lensing theory) the space is flat (Euclidean) [1,\,4].

\indent  The absence of {\it{time dilation}} is consistent not only with the SNe\,Ia observations, but also with the analysis of the variations in the light intensity of quasars by Hawkins [5], who through thorough analysis of the variability in the light intensity from near and distant quasars showed that there is no time dilation.  As we show in the following, the lensing experiments do not need any time dilation.  But time dilation is an essential part of the big-bang cosmology.

\indent  In references [1] and [3], we show that the Hubble constant, $H_0 ,$ is proportional to the average electron density, $(N_e)_{av} ,$ along the light's path from an object to the observer.  The electron densities in the coronas of galaxies are higher than the average electron densities in intergalactic space.  The galaxies therefore have intrinsic redshifts.  The electron density is highest close to the disk of the Milky Way and decreases outwards until it approaches the density in intergalactic space [2,\,3].  Consistent with expectations, SNe\,Ia data show that the Hubble constant towards an object increases not only with decreasing latitude in the Milky Way as shown in reference [3], but is also affected by the intrinsic redshifts of the lens and the source.  The lensing experiments show that quasars have a significant intrinsic redshift.

\indent  The past estimates of the Hubble constant, which were based on the big-bang cosmology, have lead to a large spread in the estimates of the Hubble constant [6].  In the present article, we will show how to interpret the gravitational lensing experiments with help of the plasma-redshift cosmology.

\indent  In section 2, we discuss briefly the aspects of the plasma-redshift cosmology relevant to lensing theory.  In section 3, we compare the distance determinations in the plasma-redshift cosmology and in the big-bang cosmology.  In section 4, we discuss the equations used in the lensing theory based on big-bang cosmology and the conventional interpretations of the lensing observations.  In section 5, we discuss the changes that have to be made to adapt the lensing equations to the plasma-redshift cosmology.  In section 6, we discuss the quantitative interpretations of the lensing observations, and in section 7, we discuss the conclusions.


\section{Plasma-redshift cosmology and gravitational lensing}


\subsection{Plasma-redshift}

\indent  In the lensing theory, we must take into account that the lensing galaxy and especially the lensed source, usually a quasar, have extended and relatively dense and hot coronas.  A hot sparse electron plasma causes a plasma redshift, $z ,$ of the photons.  When the redshift is much greater than the width of the photon, the redshift is given by (see reference [1])
\be
\frac{{d\lambda}}{{\lambda}} = \frac{{dz}}{{1+z}} = \frac{{\Phi_C}}{{2}} N_e \cdot dx\, ,
\ee
\noindent  where $\lambda$ is the wavelength, $\Phi_C = 6.6525 \cdot 10^{-25}~{\rm{cm}}^{2}$ is the cross section for the Compton scattering, and $N_e$ in ${\rm{cm}}^{-3}$ is the electron density on the stretch $dx .~$  When we integrate the left and the right side of Eq.\,(1), we get
\be
{\rm{ln}}\frac{{\lambda}}{{\lambda_0}}=  {\rm{ln}}(1+z)=  3.326 \cdot 10^{-25} \int_{R_t}^{R_h} N_e\,dx \, ,
\ee
\noindent where $R_t $ marks the lower end of the transition zone to the corona of the object and $R_h $ the outer limit of the corona.  $z $ is the redshift after penetrating the distance from $R_t~ {\rm{to}}~ R_h .~$ $\lambda_0 $ is the initial wavelength and $\lambda$ the final wavelength.  In case of an intrinsic redshift, we can replace $N_e$  by $N'_e = N_e - (N_e)_I ,$ where $(N_e)_I $ is the density of the intergalactic space.  The condition for the plasma redshift is given by (see reference [1])
\be
\lambda_{0.5} \le 318.5 \, \left( 1 + 1.3\cdot10^5 \frac{{B^2}}{{N_e}}\right)\frac{{T}}{{\sqrt{N_e}}}
\ee
\noindent  where the 50\,\% cut-off wavelength, $\lambda_{0.5} ,$ is in {\AA}-units, B the magnetic field in gauss, and T the temperature of the electron plasma in K.  This conditions means that for wavelengths of visible light the temperature must be high and the plasma density low.  Conventional laboratory plasmas do not satisfy this condition, which explains why the plasma-redshift cross section has not been discovered before.  For the cut-off wavelength $\lambda_{0.5} = 5000$ {\AA} the plasma redshift starts in the middle of the transition zone to the solar corona at about $T = 500,000$ K and at an electron density of $N_e = 10^9~{\rm{cm}}^{-3} .~$   For soft x rays, UV light, and higher magnetic fields, the cut-off moves to lower temperatures and higher densities, corresponding to the condition low in the transition zone and in the chromosphere of the Sun.

\indent   The plasma redshift often initiates the transformation of the magnetic field energy to heat [1] (see section 5.5 and Appendix B of that source).  In the Sun, the small fields are associated with spicules formation and small plasma bubbles in the transition zone.  Strong fields initiate plasma redshift and transformation of the field at greater depths and result in prominences and large flares. 

\indent  The transition zone to the corona of the Milky Way has low densities.  The plasma redshift will then start at low temperatures as soon as there are free electrons.  The light from the Milky Way and x rays from the corona and intergalactic space create the free electrons that initiate the plasma redshift in the transition zone.  On the average, all the light from the galaxies is transformed into heat, mostly with help of cosmological plasma redshift, which heats the intergalactic plasma and transforms the photon energy mostly into soft x rays.  These soft x rays return the energy to the coronas and to the transition zones of the galaxies where they ionize the medium.  The light from the galaxy with help of the plasma redshift increases the temperature to coronal temperatures.


\subsection{Plasma-redshift heating}

\indent  On a small stretch $\delta x$ of the photon's path, the photon's redshift $\delta {z}$ transfers the energy  ${\delta} {z}\,h\nu $ to the plasma, where $h\nu $ is the photons energy at that location.  The photon's energy $\delta {z} \, h \nu $ lost on the stretch $\delta x$ is absorbed locally by the plasma electrons.  The corresponding heating of the plasma is significant [1].  This energy is many orders of magnitude greater than the heating by Compton scattering.  The plasma redshift makes it therefore possible to have extended and relatively dense and hot plasma in the intergalactic space, and in the coronas of stars, quasars, and galaxies.  The hot, dense and extended coronas result in significant intrinsic redshifts of stars, quasars, galaxies, and galaxy clusters. [1].

\indent  Without the plasma-redshift heating, the sources supplying the necessary heating for the intergalactic plasma could not be found.  The intergalactic space was therefore assumed to be cold and practically empty.  The coronas of stars, quasars and galaxies were assumed to contain relatively low-density plasma.  It was also difficult to explain the great extent of some of the high-temperature H\,II regions in our Galaxy.  The density of the Milky Way's corona was assumed to correspond to only $T\,N_e  \approx 500 ~ {\rm{K\,cm}}^{-3},$ because we couldn't find enough supernovae or any other means for heating the more than a million K coronal plasma.

\indent  Other observations, such as the strength of the coronal absorption of the 6374.51 {\AA}-line from Fe\,X towards supernova 1987A, indicated about 20 times higher densities in the Milky Way's corona and temperatures exceeding a million K.  As discussed in reference [1] (see section 5.7 of that source), without plasma-redshift heating and x-ray heating from intergalactic space it was impossible to find heating sources for maintaining the high densities and high temperatures indicated by the observations.  The 20 times higher densities require 400 times as many supernovae (or other means of heating) per year to drift into the corona for balancing the cooling rate.


\subsection{Plasma-redshift leads to condensation clouds in the cornal plasma}

\indent  The plasma redshift heating is a first order process in density, while the cooling processes are of second or higher order.  The hot regions tend to get hotter and the cold regions tend to get colder.  The competition between the heating and the cooling results in large hot plasma bubbles with colder region at the surface of the bubbles and in between the bubbles [1] (see section 5.7.3 of that source).  The heat conduction counteracts and limits the rise in the temperature of the hot bubbles, while the heat conduction and heating by soft x-ray from intergalactic space often counteract the second-order emission cooling in the cold regions.

\indent   Once in a while the low temperatures in the cold regions lead to condensations, which can grow and form relatively cold plasma and clouds of hydrogen atoms and of molecular hydrogen in the coldest parts of the clouds.  In the Milky Way's corona, the hydrogen clouds can contain masses equal to many tens of millions of solar masses.  These formations together with the relatively high-density plasma will significantly increase the average density in the galactic corona.  Such clouds can also affect the structure of the lensing galaxy.

\indent  These cloud structures will have rotational velocity about equal to the rotational velocity of the plasma from which they are formed.  This plasma in turn will often have rotational velocities about equal to those of the plasma at the periphery of the transition zone.  This accounts for the observed nearly constant rotational velocity independent of radius.  As the clouds move outwards the pressure decreases and the clouds expand.  The x rays from intergalactic space can then gradually ionize the clouds and dissolve them.  The friction with intergalactic plasma slows the rotational velocities.  Clouds may occasionally reform and leak into the corona and then fall, because of their slow rotational velocities, into the transition zone of the galaxy.  These form the often-observed high-velocity clouds.  This is all consistent with observations [1] (see section 5.7 of that source). 

\indent  Such cloud structures also help explain the structures within the lens systems.  The inhomogeneity in temperature and density in the lens may cause fussy images.  These structures can also affect the magnification.   Although usually fairly stable, the clouds can grow or diminish slowly with time.


\subsection{Temperature and densities in intergalactic plasma}

\indent  We find that the intergalactic space is filled with hot plasma with an average density on the order of $(N_e)_{av} = 2.27 \cdot 10^{-4}\,(H_0/70)~{\rm{cm}}^{-3}$ and with an average temperature (per particle) on the order of $T_{av} \approx 3 \cdot 10^6\,(70/H_0)~{\rm{cm}}^{-3}.~$  The corresponding pressure counteracts the hot tail end of the pressures from the galactic coronas.

\indent  These average temperatures and densities of the intergalactic plasma explain the cosmic microwave background (CMB) [1] (see section 5.9 of that source).  In their initial reaction to my paper, some physicists did not understand how the hot plasma produces the low blackbody temperature of 2.736\,K of the CMB.  Allow me therefore to expand some details in the explanation.


\subsection{Plasma-redshift explains the CMB}

\indent  In a laboratory experiments, the temperature in the walls of the blackbody cavity determines the blackbody spectrum emitted by the cavity.  We are used to consider the temperature in the walls, or that of the particles in the walls, as determining the blackbody spectrum.  However, it is very misleading to extrapolate this to the plasma in intergalactic space, where most of the space between the particles is empty and cold.  The CMB temperature is determined by the average energy per volume, while the temperature of the plasma particles is determined by the average kinetic energy per particle.  In the plasma there is a large difference between the two quantities
 
\indent  It can be shown (Niels Bohr did that nearly 100 years ago when estimating the stopping of charged particles) that the probability for an incident particle to excite an oscillator in the medium increases with decreasing frequency of the oscillator.  In the intergalactic plasma, the highest frequencies of the Fourier harmonics of the electromagnetic field of a fast moving particle cannot reach but a tiny fraction of the empty spaces between the particles.  The energy absorbed per oscillator in a small increment $dp$ of the impact parameter $p$ is proportional to $\{\left[x\,K_1(x)\right]^2+\left[(x/\gamma)\,K_0(x)\right]^2 \} \, (dp/p), $ where $x = p\,\omega / \gamma\,v ,$ and where the function within the braces is approximately constant and equal to 1 for $x \leq 1 .~$  The Fourier harmonics, consisting of the modified Bessel function $K_1(x)$ and $K_0(x)$ of the particle's field, reach a maximum distance $ p $ from the path of the fast particle.  This maximum impact parameter is $p = \gamma\,v/\omega ,$ where $v$ is the velocity of the particle, $\gamma = 1/\sqrt{1-v^2/c^2}$ is a relativistic factor, and $\omega$ the frequency of the Fourier component.  The distance reached by the Fourier harmonic increases thus proportional to $1/\omega .~$  The charged particles in the plasma thus produce high-intensity of low-frequency harmonics that can excite any low energy oscillator.

\indent   The average volume per electron is about $17^3 \approx 5000~{\rm{cm}}^3 .~$  The impact parameter p is only about 0.225\,{\AA}  for the high-energy Fourier components produced by 300 eV electrons.  For CMB frequencies,  $\nu \geq 10^7$ Hz or $\lambda \leq 3000$ cm, the length of $p \leq 16.35 $ cm.  The CMB frequencies will permeate therefore most of the volume of space, while the high-energy Fourier components corresponding to the kinetic energy of the particles will permeate only an extremely small fraction of the volume.  In sparse plasma, there are great many low-energy CMB-frequency oscillators, consisting of all the free particles, including the free electrons, which can be excited and will then emit a continuum spectrum at CMB frequencies.  The plasma frequency, $\omega_p = 5.64\cdot 10^4 \sqrt{N_e}\approx 850$ Hz, in intergalactic space is well below the frequencies of the CMB radiation.  The particles, including the electrons, do not therefore act collectively at the frequencies of the CMB radiation.  The frequently proposed dust, grains, and whiskers are not important for explaining the CMB.

\indent  In intergalactic space, the Compton scattering determines the radius, $R_C = 6.61\cdot 10^{27}\cdot(70/H_0)~{\rm{cm}},$ of the blackbody cavity at $z = 0.5 .~$  This large blackbody cavity produces a well-defined temperature.

\indent  This underscores as shown in reference [1] that the average electromagnetic energy per volume, which determines the temperature $T_{CMB} = 2.736$ K of CMB, should not be confused with the average energy $T_p \approx 3 \cdot 10^6 $ K per particle, which determines the temperature of the plasma.  This also refutes the contention by big-bang cosmologists who usually state that the quasi-static universe cannot produce the CMB radiation, and that the CMB therefore requires and confirms the expansion theory or the big-bang cosmology. 

\indent  The excitation by ultraviolet radiation within the corresponding Str\"{o}mgren radius around stars and quasars cannot explain the high temperatures, usually more than a million K, in the coronas of galaxies and galaxy groups.  But plasma redshift with the concurrent plasma-redshift heating can increase the plasma temperature to several million K, which explains the observed excitation of highly excited ions such as Fe\,X., which requires a temperature of about 1.25 million K.


\subsection{Plasma-redshift cosmology leads to Euclidean geometry of space}

\indent  The big-bang cosmology requires dark matter for several reasons.  In the big-bang model, the intergalactic space is practically empty, and the observed density of the baryons in the universe is small, or $({\Omega}_{b} \leq 0.04), $ or at most about 4\,\% of that needed to keep the universe closed [7] (see Table 20.1 of that source).  For $H_0 = 70\,{\rm{km\,s}}^{-1}\,{\rm{Mpc}}^{-1} ,$ this value, $\Omega_b = 0.04,$ corresponds to an average density of $\rho = 3.7 \cdot 10^{-31}\, {\rm{g\,cm}}^{-3}.~$  In plasma-redshift cosmology, on the other hand, most of the baryons are in the intergalactic space, and the average density is about 1200 times higher than the above value, or about $\rho = 4.4 \cdot 10^{-28}\, {\rm{g\,cm}}^{-3}.~$  Most of the baryonic matter is in form of a hot plasma, while only a tiny fraction is in form of stars.  We can characterize it as plasma universe.

\indent  Many physicists familiar with the arguments that are used to justify the big-bang cosmology object to the high density.  They point out that consistent with Einstein's arguments, the high density, $\rho \approx 4.4 \cdot 10^{-28}\, {\rm{g\,cm}}^{-3},$ should create an observable curvature of space.  Consistent with Einstein's thinking they surmise that the gravitational laws of Newton can be extrapolated to infinity.  However, no observation justifies such an extrapolations, and quantum theory gives good reasons to believe this extrapolation is wrong.  

\indent  In spite of the high average baryonic density in intergalactic space, the plasma-redshift universe is flat.  The universe is flat because it is infinite and has nowhere to fall, apart from small-scale ripples in the gravitational potentials as those causing the gravitational lensing.  The observed flatness has been difficult to explain in the big-bang cosmology.  It has been necessary in addition to baryonic matter to introduce cold dark matter for a total of $\Omega_{m} \approx 0.3, $ which is about 8 times larger than the corresponding maximum value for baryonic matter density.  For reducing the curvature, it has also been necessary to introduce time variable $\Omega_{\Lambda} \approx 0.7 .$

\indent   Einstein, in his classical physics thinking extrapolated Newton's phenomenological equations for gravitational attraction to infinity.  Einstein was concerned about the gravitational potential, $V , $ becoming infinite when it is integrated over extreme volumes, $V = \int_0^R (G\rho/r^2)\, 4\pi r^2 \, dr \rightarrow \infty$ as R increases.  However, in the quantum mechanical world this concern is not justified.  We may not know exactly how the gravitational field interacts with matter.  However, observations show that the field must transfer information about its strength and direction to a baryonic particle.

\indent  In quantum mechanics, the time needed for such transfer of information to a particle becomes longer the smaller is the energy transferred.  (See analogous discussion of photons in reference [4].)  The strength and direction of the weak gravitational field from very distant mass cannot be transferred to a baryonic particle in the short time interval between the particle's collisions with the surrounding particles in the hot plasma universe (see the paragraphs following Eq.\,(66) in section 5.9 of [1]).

\indent The gravitational field from very distant masses therefore becomes inactive.  At the time Einstein developed his classical-physics gravitational-theory, the quantum theory had not been developed to the point that this effect could be predicted.  There is no good reason from theory or any experiment indicating that the gravitational field from very distant masses can affect a baryonic particle in a hot plasma.  Consistent with this, the curvature of space applies only to the small ripples that are subject of the gravitational lensing theory.


\subsection{Intrinsic redshifts of galaxies and quasars}

\indent  The SNe Ia data indicate that the average intrinsic redshift for latitude $b > 12^{\circ}$ in the Milky Way is on the order of $\delta z'_{MW}= 0.00095 $ [2] with a definite increase with decreasing latitude [3].

\indent  The lensing galaxy or the lensing group of galaxies may be surrounded by dense galactic corona.  The intrinsic redshift, $\delta z'_L , $ can be expected to be as large or larger than that of the Milky Way.  The redshift, $z'_L , $ determining the distance to the lensing galaxy is then given by $z'_L = z_L - \delta z'_{MW} - \delta z'_L ,$ where $z_L $ is the observed redshift.

\indent   The source, which is often a quasar, may have exceptionally large intrinsic redshift, $\delta z'_S .~$  The light path to the source grazes the lens, which results in an intrinsic redshift $\delta z'_{LL} $ in addition to $ \delta z'_{L} .~$  For obtaining the redshift $z'_S$ that determines the distance to the source, we must replace the observed redshift $z_S $ by $z'_S = z_S - \delta z'_S - \delta z'_{LL}-\delta z'_{L} -  \delta z'_{MW} .~$  It therefore will be necessary to modify significantly the present lensing theory.


\section{Variation of the distances with the redshift z}


\subsection{Distances in plasma-redshift cosmology}

In plasma redshift, the incremental redshift is given by Eq.\,(1) and (2).  It is proportional to the electron density, $N_e ,$ along the path, $dx ,$ of the photon.  From Eq.\,(2) it follows, as shown in reference [1,\,3] (see in particular Eq.\,(48) of [1], or Eq.\,(4) of reference [3]), that the distance $D$ to an object is then given by
\be
D = \frac{{c}}{{H_0}}\,{\rm{\ln}}(1+z)= \frac{{3000}}{{h}}\,{\rm{\ln}}(1+z)\,~{\rm{Mpc}},
\ee
\noindent  where $c=3 \cdot 10^5 ~ {\rm{km\,s}}^{-1}$ is the velocity of light, $h=H_0/100 ,$ and $c / H_0 $ is the Hubble length.  The Hubble constant, $H_0 ,$ is
\be
 H_0 = 3.0764\cdot 10^5 \,(N_e)_{av} ~ {\rm{km\,s}}^{-1}\,{\rm{Mpc}}^{-1}
\ee
\noindent  and $(N_e)_{av}$ in ${\rm{cm}}^{-3}$ is the average electron number density in the plasma along the distance $D.~$  For example, if the average electron density along the distance $D$ is $(N_e)_{av} = 2.34\cdot 10^{-4}~{\rm{cm}}^{-3}$, the Hubble constant is $H_0 = 3.0764\cdot 10^5 \cdot 2.34\cdot 10^{-4}= 72~{\rm{km\,s}}^{-1}\,{\rm{Mpc}}^{-1}.~$  The value of $H_0 =  72~{\rm{km\,s}}^{-1}\,{\rm{Mpc}}^{-1},$ is equal to the average determined in the Hubble Space Telescope Key Project [6].  The value of $(N_e)_{av} = 2.34\cdot 10^{-4}~{\rm{cm}}^{-3}$ includes the intrinsic values of the Milky Way and the lensing galaxy or galaxy cluster.  In intergalactic space, the average density is about $(N_e)_{av} \approx 1.95\cdot 10^{-4}~{\rm{cm}}^{-3}$ and the corresponding Hubble constant is $H_0 \approx 60~{\rm{km\,s}}^{-1}\,{\rm{Mpc}}^{-1} $ [3].  In the lensing experiments, the light path grazes the lens.  The path close to the lensing galaxy or the lensing galaxy cluster will then often result in a relatively large average electron density and therefore a large Hubble constant.


\begin{figure}[t]
\centering
\includegraphics[scale=.5]{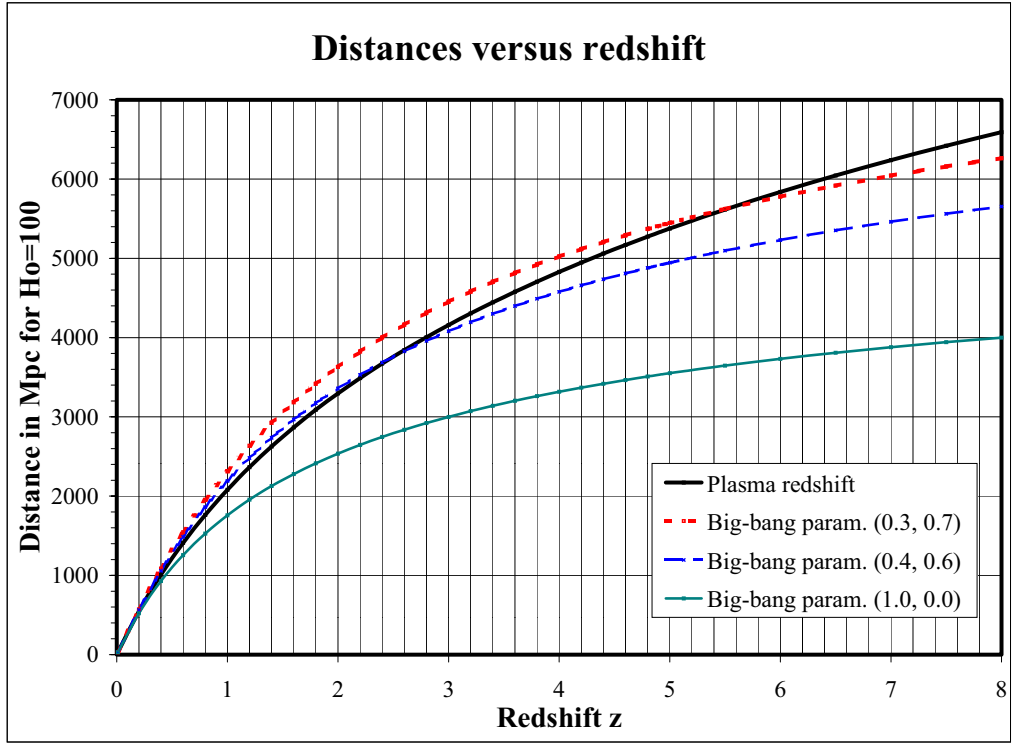}
\caption{The ordinate gives the distance in Mpc for a Hubble constant $H_0 = 100\,{\rm{km\,s}}^{-1}\,{\rm{Mpc}}^{-1}$ versus the redshift, $z$ from 0 to 8 on the abscissa.  The solid upper curve gives the distance in a cosmology based on the plasma redshift, while the small-dashed curve, the long dashed curve and the solid lower curve give the distances based on the flat big-bang cosmology when the cosmological parameters $(\Omega_m,\,\Omega_{\Lambda}) = (0.3,\,0.7), \,  (0.4,\,0.6) , \,{\rm{and}}\, (1.0,\,0.0) ~$ respectively, and when $H_0 = 100\,{\rm{km\,s}}^{-1}\,{\rm{Mpc}}^{-1}.~$  It can been seen that we can adjust the cosmological parameters in the big-bang cosmology to give a fairly good fit (the small dashed and the long dashed curves) to the plasma redshift theory.}
\vspace{2mm}
\end{figure}


\subsection{Distances in big-bang cosmology}

In the big-bang cosmology the distance-redshift relation is more complicated, because in addition to the redshift $z$ and the Hubble constant, $H_0 ,$ the distance depends on at least three parameters: the mass density parameter $\Omega_m ,$ the expansion parameter $\Omega_{\Lambda},$ and the curvature parameter $\Omega_k .~$  The dark matter parameter is usually included in $\Omega_m .~$  The expression for the comoving distance is (see reference [7,\,8], or Eq.\,(B1) of [9])  
\be
D = \frac{{c}}{{H_0}}\,\frac{{1}}{{ |\Omega_k|^{1/2} }}\,{\rm{sinn}}\left[ \int_0^z \frac{{ |\Omega_k|^{1/2} \,dz'}}{{\,\left[ (1+z')^2(1+{\Omega}_m \,z') - z'(2+z') {\Omega_{\Lambda}} \right]^{1/2} \,}}\right],
\ee
where $c / H_0$ is the Hubble length, $\Omega_k = 1 - \Omega_m -\Omega_{\Lambda}$ is the curvature parameter, and $\Omega_m = 8 \pi G \,\rho/(3H_0^2)$ is the mass density parameter, $G = 6.672 \cdot 10^{-8}~{\rm{cm}}^3 \, {\rm{s}}^{-2} \, {\rm{g}}^{-1} $ is the gravitational constant, and $\rho~{\rm{in~g\,cm}}^{-3}$ is the average mass density in the universe.  $\Omega_{\Lambda} = \Lambda/(3H_0^2)$ is the expansions parameter, and $\Lambda$ the cosmological constant.  Often used values for these parameters in flat space are $(\Omega_m , \, \Omega_{\Lambda}) = (0.3, \, 0.7).~$  In flat space, we must have that $\Omega_k = 0 ,$ or $\Omega_m \, + \, \Omega_{\Lambda} = 1.~$  The function sinn\,(x) is equal to sinh(x) for $\Omega_k > 0$, equal to x for $\Omega_k = 0 ,$ and equal to sin\,(x) for $\Omega_k < 0 .~$  These parameters are usually adjusted to fit the observations.

\indent  In the limit of $\Omega_k = \Omega_m = \Omega_{\Lambda} = 0 ,$ Eq.\,(6) becomes identical to Eq.\,(4).  However, in the big-bang cosmology, the assumed time dilation, the cosmological parameters for density, expansion, and curvature affect the estimates of distances and the interpretations of many observations.  For matching their predictions with the observations, the big-bang cosmologists adjust all the cosmological parameters in Eq.\,(6) together with the Hubble constant $H_0 .$  
  
\indent Fig.\,1 illustrates how the distances in flat space vary with the redshift.  It indicates that the parameters, $\Omega_m,\,\Omega_{\Lambda},~{\rm{and}}~\Omega_k $ can be adjusted to give a fairly good fit to the plasma-redshift theory.  For $(\Omega_m,\,\Omega_{\Lambda}) = (0.3,\,0.7)$ and flat space the maximum deviation is about $\delta D = 11.4\,\%$ at $z = 1.2 .~$  This corresponds to change in absolute magnitude of $\delta M = 0.23 ,$ which is about equal to or less than the accuracy in the measurement of the magnitude of the SNe\,Ia.


\subsection{Angular diameter distances}

When using the lensing theory in the big-bang cosmology, it is customary to use the so-called comoving angular diameter distances $^{\rm{a}}\!D,$ which are obtained by dividing the familiar comoving distances in Eq.\,(6) by  $(1 + z) ,$ that is,  $^{\rm{a}}\!D,$ is related to the actual distance, $D,$ as
\be
^{\rm{a}}\!D = \frac{{D}}{{(1+z)}} ,
\ee
\noindent  where $D$ is defined by Eq.\,(6).  The angular distances are simpler to use when developing and interpreting the lensing equations.

\indent  Eq.\,(7), applies to the distances $D$ and the redshifts $z$ from the observer to the source and from the observer to the lens.  For distances from the lens to the source the relation is
\be
^{\rm{a}}\!D_{LS} = \frac{{1}}{{(1+z_S)}} \left[D_S \sqrt{1+(D_L / D_H)^2\,\Omega_k}~-~D_L \sqrt{1+(D_S / D_H)^2\,\Omega_k}\, \right]  ,
\ee
\indent  We may also use the dimensionless quantities
\be
 d = \frac{{H_0}}{{c}} \, ^{\rm{a}}\!D .
\ee 
\indent When using the plasma-redshift cosmology in the lensing theory, there is no expansion and no need for the so-called angular distances.  Eq.\,(4) gives the distances.


\subsection{Luminosity distances}

In the big-bang cosmology, the luminosity distance is slightly greater and given by $D_{lu}=D(1+z).~$  This corresponds to the light intensity $I$ decreasing with distance as

\be 
I = \frac{{I_0}}{{D_{lu}^2}} = \frac{{I_0}}{{D^2(1+z)^2}}.
\ee
\noindent  One factor, $1/(1+z),$ is due to reduction in each photon's energy caused by the redshift.  The other factor $1/(1+z)$ takes into account the time dilation $(1+z),$ which causes reduced rate of photon emission as seen by the observer.

\indent  In the plasma-redshift cosmology, the luminosity distance is still greater and given by $D_{lu} = D(1+z)^{3/2}.~$  This corresponds to the bolometric light intensity $I$ decreasing with distance as

\be 
I = \frac{{I_0}}{{D_{lu}^2}} = \frac{{I_0}}{{D^2(1+z)^3}}.
\ee
\noindent  One factor, $1/(1+z),$ accounts for the reduction in each photon's energy caused by the redshift, while the factor $1/(1+z)^2$ accounts for the Compton scattering, which removes the photon out of line of sight.  When observing a distant star or a quasar, most of the scattered light from the coronal electrons would be observed as coming from the star or the quasar.  It is usually important to distinguish intrinsic redshift from the redshift caused by interaction with electrons in space along the line of sight.  The cross section for intensity reduction caused by the Compton scattering is exactly twice that of the plasma redshift.  In the plasma redshift model there is no time dilation, as the universe is quasi-static.

\indent  There is a clear difference between the predictions of Eqs.\,(10) and (11).  The absolute magnitudes of the SNe\,Ia increases with the width of the light curve.  This causes a significant Malmquist bias.  When this is taken into account, the predicted magnitude-redshift relation in the plasma-redshift cosmology matched the observations [2].  We consider this match as confirming the plasma-redshift cosmology over the conventional big-bang cosmology.  In reference [2], we corrected the absolute magnitude of the SNe\,Ia for the Malmquist bias based on the reported observations


\section{Lensing in the big-bang cosmology}


\subsection{Lensing equations in the big-bang cosmology}

Gravitational lensing experiments are conventionally explained on the basis of the big-bang cosmology.  Table 1 shows that the estimates of the Hubble constant are in the range of $H_0 \approx 48~{\rm{to}}~ 78\, {\rm{km\,s}}^{-1}\,{\rm{Mpc}}^{-1}.~$  The average, $H_0 \approx 62.3\, {\rm{km\,s}}^{-1}\,{\rm{Mpc}}^{-1},$ is lower than the average $H_0 \approx 72\pm 8 ~{\rm{km\,s}}^{-1}\,{\rm{Mpc}}^{-1}$ obtained in the Hubble Space Telescope Key Project [6], which combined the results of different methods (not including lensing methods).

\indent  The SNe\,Ia researchers have usually selected supernovae that are nearly free of peculiar absorptions.  This reduces the interference of dense plasmas and the Hubble constant.  When comparing the SNe\,Ia data with the predictions of the plasma-redshift cosmology [2], we tried to eliminate most of the intrinsic redshifts.  This resulted in further reduction in the estimates of the Hubble constant to $H_0 \approx 60 ,$ which corresponds to the average electron densities in intergalactic space. 

\indent  In case of lensing experiments little or no effort has been made to reduce the intrinsic redshifts.  Only a few lensing systems are available.  Also, the light path has to penetrate the entire length of the relatively dense corona of the lensing galaxy or galaxy group.  The estimates of the average value of $H_0,$ which depends on the average electron density along each distance may skew the estimates.  When we compare the observed data with the predictions of both the big-bang and the plasma-redshift cosmology and disregard intrinsic redshifts, we should expect relatively large values for the Hubble constant.


\indent  The conventional gravitational lensing models are described well by Peebles [7], Kochanek, Schnider and Wambsganss [9], Narayan and Bartelmann [10], Schechter [11], and Kochanek and Schechter [12].  In these models it is usually necessary to make some assumptions about the gravitational mass distribution and the gravitational environment including distribution of dark masses.  It is a common practice in the big-bang cosmology to surmise that the galaxies and quasars have no intrinsic redshifts, except those caused by the Doppler shifts.   

\indent  For facilitating the discussion, we borrow from Schechter [11] (see Eq.\,(3.3) of that source) and Cochanek et al. [9] (see in particular Eq.\,(B93) of that source) the time delay function
\be
\tau (\theta) \, H_0  = (1+z_L)  \, \frac {{\, d_L \, d_S }}{{ d_{LS} }}\left[ \frac{{1}}{{2}} (\vec{\theta} - \vec{\beta})^2 \, -\, {\psi}(\vec{\theta})\right ] = \frac{{H_0}}{{c}}\, \frac {{\, D_L \, D_S }}{{ D_{LS} }}\left[ \frac{{1}}{{2}} (\vec{\theta} - \vec{\beta})^2 \, -\, {\psi}(\vec{\theta})\right ],
\ee
\noindent where $\tau$ is the time delay caused by the gravitational potential along the ray and the increased path caused by the bending of the ray relative to the unperturbed ray from the source.  $H_0$ is the Hubble constant, $z_{L}$ is the redshift of the lens galaxy, and the quantities $d_L = (H_0/c) \,D_L / (1+z_{L}), ~ d_S = (H_0/c)\,D_S / (1+z_S), ~ {\rm{and}} ~ d_{LS} = (H_0/c)\, D_{LS} / (1+z_S),$ are the dimensionless representations of the angular distances $^{\rm{a}}\!D_L$ to the lens, $^{\rm{a}}\!D_S$ to the source, and the distance $^{\rm{a}}\!D_{LS}$ from the lens to the source, respectively.  ${\vec{\beta}} $ is the angle between the lens galaxy and the source, while  ${\vec{\theta}}$ is the angle between the lens galaxy and the image of the source.  $( {\vec{\theta}} - {\vec{\beta}} )= \alpha = \hat{\alpha}(D_{LS} / D_S) $ is therefore the angle between the source and its image, while $\hat{\alpha}= 4GM(\xi)/(c^2 \xi) $ is the angle between the source and the image as seen by an observer at the lens.  This angle $\hat{\alpha}$ is equal to the gravitational deflection at the lens.  The three dimensional mass distribution in the lensing galaxy is projected onto the lens plane, which is perpendicular to the line of sight.  $M(\xi)$ represents the projected mass distribution on to the lens plane.  It is the lens mass inside the radius $\xi$ from the center of the lens in this plane.  $G$ is the gravitational constant.  ${\psi}(\vec{\theta})$ is the two-dimensional effective potential.  At each position the gradient of ${\psi}(\vec{\theta})$ with respect to $\theta $ is the deflection angle.  We have $ \vec{\bigtriangledown}{\psi}(\vec{\theta}) = (\vec{\theta} - \vec{\beta}) =\vec{\alpha}.~$


\subsection{Determination of $H_0 $ in the big-bang cosmology}

We can have more than one image of a source lensed by a galaxy.  For each image, we get equations analogous to Eq.\,(12).  The photon's travel-time ${\tau}_{B} $ from the image $B$ will usually differ from the travel time ${\tau}_{A} $ from the image $A .~$  The time delay difference, $({\tau}_{B}-{\tau}_{A}) ,$ can sometimes be measured.  For example, a peculiar variation in light intensity $I$ from the source may occur first in the image $A$ and subsequently, maybe 100 days later, in image $B$.  From the distances in the lensing geometries, and from the velocity of light (which is affected by the gravitational potentials along the photon's path) the difference, $({\tau}_{B}-{\tau}_{A}) ,$ in light's travel times can be calculated.  This calculated time delay, $({\tau}_{B}-{\tau}_{A}) ,$ can then be compared with the observed time delay and used to calibrate the distance scale.  The calculated time delay can then be used to determine also the Hubble constant $H_0 .$

\indent  The estimated value of the Hubble constant depends on the model used for the lens.  A useful reference model for the lens is the {\it{singular isothermal sphere}} discussed by Kochanek [9] (see in particular section B3.2 of that source).  This model assumes that the corona of the lensing galaxies is nearly isothermal and that the density, $\rho ,$ decreases outwards about proportional to $R^{-2} ,$ where $R$ is the radius from the center of the lensing galaxy. This model for a galaxy fits many observations.  Kochanek shows that in this case Eq.\,(12) leads to the following equation [9] (see Eq. B.94 of that source)
\be
( {\tau}_{B}-{\tau}_{A} )\, H_0 =  (1+z_L) \, \frac {{\, d_L \, d_S }}{{ d_{LS} }}\, \frac{{1}}{{2}} \left[ |{\vec{\theta}}_A|^2 - |{\vec{\theta}}_B|^2 \right] = \frac{{H_0}}{{c}} \, \frac {{ D_L \, D_S }}{{ D_{LS} }}\, \frac{{1}}{{2}} \left[ |{\vec{\theta}}_A|^2 - |{\vec{\theta}}_B|^2 \right]  \,  ,
\ee
where $\theta_A$ and $\theta_B$ are the observed angular displacements of the images A and B from the center of the lensing galaxy.

\indent  For more elaborate models of the lens, we obtain (see Eq.\,(4.3) of reference [11] and Eq.\,(B.97) of reference [9]) that
\be
( {\tau}_{B}-{\tau}_{A} ) \, H_0 = \frac{{H_0}}{{c}} \, \frac {{ D_L \, D_S }}{{ D_{LS} }}\, \left[ |{\vec{\theta}}_A|^2 - |{\vec{\theta}}_B|^2 \right] \,  f(\kappa,\, \eta, \, {\theta}_A, \, {\beta}_A, \, {\theta}_B, \, {\beta}_B) ,
\ee
\noindent  where the structure function $f(\kappa,\, \eta, \, {\theta}_A, \, {\beta}_A, \, {\theta}_B, \, {\beta}_B) = 1/2 $ for the {\it{singular isothermal sphere}} model of the lensing galaxy.  More generally, this structure function varies with the density and spatial distribution of the density in the corona of the lensing galaxy, and with the lensing angels $\theta$ and $\beta$ for each image A and B.  In reference [9] (see in particular Eq.\,(B.97) of that source) Kochanek et al. find that if the density varies with the radius from the center of the lens but is isotropic, the structure function takes the form


\begin{table}[h]
\centering
{\bf{Table 1.}}  The Hubble constant $H_0$ as determined by gravitational lensing and big-bang model

\vspace{2mm}

\begin{tabular}{llllllll}
\hline
~~System   &${\rm{N_{im}}}$&$\Delta \tau~({\rm{days}})$& $z_{S}$ & $z_L $ & $H_0$ & Ref \\
\hline \hline
HE1104-1805 &~2   &$310\pm{18}$ &2.319 &0.729 &$48\pm{4};~62\pm{4}$                       &[13,\,14] \\
HE1104-1805 &     &161$\pm$7    &      &      &$>75$                                      &[15,\,16] \\
PG1115+080  &~4   &~25$\pm$2    &1.722 &0.311 &$49_{-11}^{+6};~56_{-11}^{+12}$            &[17,\,18,\,19,\,20] \\
SBS1520+530 &~2   &130$\pm$3    &1.86  &0.717 &$51\pm 9;\,63\pm 9$                        &[21] \\
B1600+434   &~2   &~51$\pm$2    &1.59  &0.41  &$52_{-8}^{+8};~60_{-19}^{+30};~78_{-27}^{+28}$&[22,\,23] \\
HE2149-2745 &~2   &103$\pm$12   &2.03  &0.495 &$66\pm 9 $                                 &[24] \\
RXJ0911+0551&~4   &146$\pm$4    &2.80  &0.77  &$48\pm 3;~71\pm 9 $                        &[25,\,26] \\
Q0957+561   &~2   &417$\pm$3    &1.41  &0.36  &$66\pm13$                                  &[27,\,28] \\
B1608+656   &~4   &~77 $\pm$2   &1.394 &0.630 &$63\pm{15};~75_{-7}^{+8}$                  &[29,\,30,\,31] \\
B0218+357   &~2   &~10.5$\pm$0.2&0.96  &0.685 &$69_{-19}^{+13};~73\pm 8;\,75_{-23}^{+17}$ &[32,\,33,\,34] \\
B1422+231   &~4   &~(8$\pm$3)   &3.62  &0.33  &$\approx\,64~{\rm{to}}~\approx 75$         &[35] \\
FBQ0951+2635&~2   &(16$\pm$2)   &1.246 &0.24  &$60_{-7}^{+9},~63_{-7}^{+9}$               &[36] \\
\hline
\end{tabular}
\end{table}

\be
f(\kappa,\, \eta, \, {\theta}_A, \, {\beta}_A, \, {\theta}_B, \, {\beta}_B) = \left[ {1 - {\langle{\kappa}\rangle} - \frac{{ 1-\eta ( \langle {\kappa}\rangle ) }}{{ 12 }}\,\left( \frac{{ {\delta \theta} }}{{ {\langle{\theta}\rangle} }}\right)^2 \, + \, O \left( \left( \frac{{ \delta \theta }}{{ {\langle{\theta}\rangle} }}  \right)^4  \right)} \right]
\ee
\noindent  where it is assumed that the surface density $ \kappa (\theta) = \langle {\Sigma (D_L \theta)} \rangle /\Sigma_{cr}  $ in the annulus $\theta_B < \theta < \theta_A $ can be approximated by $\kappa(\theta) \propto {\theta}^{1-\eta}.~$  The annulus $\theta_B < \theta < \theta_A$ is formed by the two circles, which are drawn around the lens through the images A and B with the lens at the center.  $\Sigma$ is the two-dimensional mass distribution obtained by projecting the three-dimensional mass distribution of the lens onto the lens plane, which is perpendicular to the line of sight to the lens.  $\Sigma_{cr} = \left[c^2/(4\pi G)\right]\left[D_S/(D_L D_{LS})\right] $ is the critical surface density.  Other forms (including elliptic and anisotropic forms) of the structure functions can be developed, when specific information about the lens and its environment are available. 

\indent   Table 1 lists some of the parameters for the 11 lensing systems that have been used to determine the value of $H_0 .~$ The first column gives the name of the lensing system; the second gives the number of source images; the third gives the observed time delay (in days) in the intensity of one image compared with another; the fourth and fifth columns give the source and lens redshifts, the sixth column gives the estimated value of $H_0 $ assuming a flat cosmology where $\Omega_m =0.3$ and $\Omega_{\Lambda}=0.7 .~$  The seventh column gives the references.

\indent  The two different time delays for the system HE1104-1805, $\Delta\tau\approx 310~{\rm{and}}~\Delta \tau \approx 161$ days, could not both be right.  Both results are based on thorough investigations.  Interference from lensing objects can mislead and explain the difference.  In the following, we surmise that the time delay is $\Delta \tau \approx 310 $ days.

\indent  The lens of the system B1608+656 belongs to a group of galaxies [31].  This lens is then surrounded by relatively dense and extensive hydrogen plasma, which will result in large in large intrinsic redshift $\delta z'_L $ and a large $ z_L$-value.

\indent  Also the lens Q0957+561 resides in a cluster of galaxies, and is therefore likely to have relatively large $\delta z'_L ,$ and large $z_L.~$  For the system He2149-2745, the $z_L = 0.495$-value is uncertain and may be $z_L = 0.489$ instead.  The high values of $H_0$ in table 1 may therefore be caused by rather high $z_L$-values.  

\indent  The source in the system B0218+357 is special.  It is a BL Lac object.  The jet from this BL Lac source appears to be pointing at us, and we are then looking at the head of the jet with a smaller intrinsic shift than a regular quasar.  The shift may even be a small intrinsic blue shift rather than a small redshift.

\indent  For estimating $H_0 ,$ the big-bang cosmology is assumed valid.  Usually it is assumed that the curvature parameter ${\Omega}_k = 0 ,$ the density parameter $\Omega_m = 0.3 ,$ and the expansion parameter ${\Omega}_{\Lambda} = 0.7 .~$   In Table 1 the $H_0$-values are adjusted to this cosmology.  For example, when the Cohen et al. [34] in case of B0218+357 assumed flat cosmology based on $\Omega_m = 1,$ their reported $H_0 = 71_{-23}^{+17} $ was increased by 6\,\% to $H_0 = 75_{-23}^{+17} $ for adjusting the values to $(\Omega_m,\,\Omega_{\Lambda})= (0.3,\,0,7)$-cosmology.


\section{Lensing in the plasma-redshift cosmology}

The gravitational deflection of light depends on the velocity of light and the warping of space in accordance with the general theory of relativity (GTR).  This deflection is independent of the frequency of light and the weight or weightlessness of the photon [4].  The quantum mechanically modified GTR [4] therefore predicts identical lensing of light to that of the classical GTR.  

\indent  The basic gravitational lensing theory in the cosmology based on the big-bang model is the same, therefore, as that in the cosmology based on plasma redshift and the quantum mechanically modified theory of GTR.  However, when applying the Eqs.\,(12) to (15) above, we must use the forms with comoving distances, $D ,$ and not the angular distances, $^{\rm{a}}\!D_{L} = D_{L}/(1+z_L) ,$  or the dimensionless distances, 
$d_L =(c/H_0)\, ^{\rm{a}}\!D_{L}.~$  In the lensing theory based on plasma redshift, we must:   

\begin{enumerate}
\item	Replace the comoving distances given by Eq.\,(6) in the big-bang cosmology with the distances in the plasma redshift cosmology, which are given by Eq.\,(4).
\item	Take into account the intrinsic redshifts of the different objects, especially the large intrinsic redshifts of the quasars. 
\end{enumerate}

\indent  The average intrinsic redshift corrections $\delta z'_{MW} \approx 0.00095 $ along the path close to the Milky Way, the intrinsic redshift $\delta z'_{L}$ along the path close to the lens, the intrinsic redshift $ \delta z'_{LL}$ along the path grazing the lens, and intrinsic redshift $ \delta z'_{S}$ along the path close to the source can all be significant, especially the intrinsic redshift $\delta z'_{S}$ of the source, which is usually a quasar.

\indent  For estimating the distances to the source, the observed source redshift $z_S $ must then be replaced by
\be
z'_S = z_S - \delta z'_{MW} - \delta z'_{L} - \delta z'_{LL}-\delta z'_S ~;
\ee
\noindent  the observed redshift $z_{LS}$ must be replaced by
\be
z'_{LS}=z_{LS} - \delta z'_{LL}-\delta z'_S ~. 
\ee
\noindent  and the observed lens redshift $z_L $ must be replaced by 
\be
z'_L = z_L - \delta z'_{MW} - \delta z'_{L}~;
\ee
\indent  When transforming the equations in big-bang cosmology to the corresponding equations in plasma-redshift cosmology, we can in Eqs.\,(12) and (14), which are valid in flat big-bang cosmology, set the factor in front of the brackets equal to
\be (1+z_L) \, \frac{{ d_L \, d_S }}{{ d_{LS} }} = (1+z_L) \, \frac{{\left[ \frac{{H_0}}{{c}} \, \frac{{ D_L }}{{ (1+z_L) }}\right] \cdot \left[ \frac{{H_0}}{{c}} \, \frac{{D_S}}{{(1+z_S)}}\right] }} {{\left[ \frac{{H_0}}{{c}} \, \frac{{ D_{LS} }}{{ (1+z_{S} ) }}\right] }} = \frac{{H_0}}{{c}} \, \frac{{ D_L \, D_S }}{{ D_{LS} }}~ ,  
\ee
\noindent  where both the left and right side are valid in the big-bang cosmology.  We can equate the right side of Eq.\,(19) with the corresponding equation in the plasma-redshift cosmology.  When using Eq.\,(4) for the distances, we get from Eq.\,(19) that
\be
 (1+z_L) \, \frac{{ d_L \, d_S }}{{ d_{LS} }} \Longrightarrow \frac{{H_0}}{{c}} \, \frac{{ D_L \, D_S }}{{  D_{LS} }} =  \frac{{ {\rm{ln}}(1+z'_L) \, {\rm{ln}}(1+z'_S) }}{{ {\rm{ln}}(1+z'_{S}) - {\rm{ln}}(1+z'_{L}) }} ~.
\ee
\noindent  When we insert Eqs.\,(16), (17), and (18) into Eq.\,(20), we get
\be
 \frac{{H_0}}{{c}} \, \frac{{ D_L \, D_S }}{{ D_{LS} }} = \frac{{ {\rm{ln}}(1+z_L - \delta z'_{MW} - \delta z'_{L}) \, {\rm{ln}}(1+z_S - \delta z'_{MW} - \delta z'_{L} - \delta z'_{LL}-\delta z'_S) }}{{ {\rm{ln}}(1+z_S - \delta z'_{MW} - \delta z'_{L} - \delta z'_{LL}-\delta z'_S) - {\rm{ln}}(1+z_L - \delta z'_{MW} - \delta z'_{L}) }}
\ee
\indent  The scaled gravitational potential term,  ${\psi}(\vec{\theta})$ inside the brackets of Eq.\,(12) depends on the observed angles, and is therefore unchanged.  However, the change in distances will affect the assumed mass and mass distribution.  As Eq.\,(13) indicates, such changes will not affect the value of $(\tau_B - \tau_A).~$  If the intrinsic redshifts were insignificant the changes in the mass distribution would be small, because Eq.\,(6) for $(\Omega_m,\,\Omega_{\Lambda})= (0.3,\,0,7)$ is fairly close to the plasma-redshift distances given by Eq.\,(4), as illustrated in Fig.\,1.   We get then 
\be
{\psi}(\vec{\theta})= \frac {{ D_{LS} }}{{ D_{S} }}\, 4GM(\xi)/(c^2 \xi) \Longrightarrow {\psi}'(\vec{\theta})= \frac {{ D'_{LS} }}{{ D'_{S} }}\, 4GM'({\xi}')\,/\,(c^2\, {\xi}') .
\ee
\noindent  In this equation we have that the observed quantity ${\psi}(\vec{\theta})= {\psi}'(\vec{\theta}) ;$ but the different quantities on the right hand sight of the equations differ.  Roughly, $M(\xi) \simeq M'(\xi') / (1+z_L)$ and $\xi \simeq \xi' / (1+z_L) .$ 

\indent  In plasma-redshift cosmology, Eq.\,(14) takes the form
\be
 ( {\tau}_{A}-{\tau}_{B} ) \, H_0 = \frac{{ {\rm{ln}}(1+z'_L ) \, {\rm{ln}}(1+z'_S ) }}{{ {\rm{ln}}(1+z'_{S}) - {\rm{ln}}(1+z'_{L}) }} \, \left[ |{\vec{\theta}}_A|^2 - |{\vec{\theta}}_B|^2 \right] \,  f(\kappa,\, \eta, \, {\theta}_A, \, {\beta}_A, \, {\theta}_B, \, {\beta}_B) ,
\ee
\noindent  where $z'_S ,$  $z'_{LS} ,$ and  $z'_{L}$ are given by Eqs.\,(16), (17), and (18), respectively; and where the structure function $f(\kappa,\, \eta, \, {\theta}_A, \, {\beta}_A, \, {\theta}_B, \, {\beta}_B) = 1/2 $ for the {\it{singular isothermal sphere}} model of the lensing galaxy.  Eq.\,(23) is then analogous to Eq.\,(13) for the {\it{singular isothermal sphere}} model in the big-bang cosmology.


\section{Comparing the different theories with observations }

In the Hubble Space Telescope Key Project [6], the Hubble constant is found to be $H_0 =  72~\pm{8}~{\rm{km\,s}}^{-1}\,{\rm{Mpc}}^{-1}.~$  As pointed out in references [2] and [3], the plasma redshift theory leads to much simpler magnitude-redshift relation than that in the big-bang cosmology.  There is no time dilation, and no cosmological parameters, $\Omega_m ,~\Omega_{\Lambda} ,~{\rm{and}}~ \Omega_k ,$  for adjusting the theory to the observations.  The Hubble constant, which is no longer an expansion parameter, is replaced by Eq.\,(5): $H_0 = 3.0764\cdot 10^5 \,(N_e)_{av} ~ {\rm{km\,s}}^{-1}\,{\rm{Mpc}}^{-1}.~$  Within the coronas of galaxies, the electron densities are often on the order of or exceed $N_e = 10^{-2}~{\rm{cm}}^{-3},$ while the average electron density in intergalactic space is $N_e \approx 2 \cdot 10^{-4}~{\rm{cm}}^{-3}.~$  The Milky Way, lens galaxies, galaxy clusters, and quasars often have significant intrinsic redshifts.


\begin{table}[h]
\centering
{\bf{Table 2}} \, \, Hubble constant $H_0$ from lensing
\vspace{2mm}

\begin{tabular}{llllllll}
\hline \hline
& $~F_{Bb}$&$~F_{Pl}$&$({H_{0}})_{Bb}$&$(H_0)_{Pl}$&$(H_0)_{Pl}$&$(H_0)_{Pl} $\\
~~~System &&& &$z'_Q=0$&$z'_Q=(z_S-z_L)/3$&$z'_Q=(z_S-z_L)/2$\\
\hline 
HE1104-1805 &~1.1227&~1.0073&~~55.0&~~49.4 &~~57.5      &~65.6 \\
PG1115+080  &~0.3894&~0.3712&~~52.5&~~50.0 &~~54.8      &~59.5 \\
SBS1520+530 &~1.2270&~1.1133&~~57.0&~`51.7 &~~62.1      &~72.4 \\
B1600+434   &~0.5700&~0.5377&~~63.3&~~59.7 &~~67.7      &~75.7 \\
HE2149-2745 &~0.6812&~0.6310&~~66.0&~~61.1 &~~69.0      &~76.8 \\
RXJ0911+0551&~1.1263&~0.9977&~~59.5&~~52.7 &~~60.5      &~68.2 \\
Q0957+561   &~0.4964&~0.4727&~~66.0&~~62.9 &~~71.2      &~79.4 \\
B1608+656   &~1.1976&~1.1096&~~69.0&~~63.9 &~~78.7      &~93.5 \\
B0218+357   &~2.4790&~2.3225&~~72.3&~~67.7 &~~92.1(67.7)&~116.4(67.7) \\
B1422+231   &~0.3753&~0.3505&~~64.0&~~59.8 &~~62.9      &~65.9 \\
FBQ0951+2635&~0.3024&~0.2930&~~60.0&~~58.1 &~~63.9      &~69.6 \\
\hline
Average $H_0$&      &       &~~62.2&~~57.9 &~~67.3(65.1)&~76.6(72.2) \\
\hline \hline
\end{tabular}
\end{table}

\indent  From our position in the Milky Way, the intrinsic redshifts in the Milky Way's corona are $\delta z'_{MW} \approx 0.00095 $ for latitudes $b > 12^{\circ} .~$  Nevertheless, it is clear from the supernova-observations, as shown in reference [3], that the Hubble constant increases with the object's decreasing latitude in the Milky Way even for $b > 12^{\circ} .~$  In the lensing experiments, the light path from the source grazes the lensing object.  The average electron density along the path of the light grazing a galaxy is higher than in intergalactic space.  The light-path may graze the bulge of the lensing galaxy, where the coronal density is highest.  The spaces between galaxies in galaxy clusters often have relatively high electron densities.  We expect therefore the average Hubble constant in lensing experiments to be higher than that obtained in the SNe\,Ia experiments.  From Table 2 we see that this expectation contradicts the average value of $H_0 \approx 57.9 $ when $ \delta z'_Q = 0 $.

\indent  Most of the lensed sources are quasars, which may have a significant intrinsic redshift.  We suspect that the lower than expected value of $H_0$ in the lensing experiments may be due to large intrinsic redshift of the lensed quasar.  The determination of the Hubble constant, $H_0, $ from lensing experiments depends on the ratio $D_L \, D_S/D_{LS}, $ which depends on the intrinsic redshifts as shown in Eqs.\,(20) and (21).  In the first approximation, we assume that $\delta z'_{MW} $ and $\delta z'_{L} $ are small.  When we then in Eqs.\,(15) and (16) set $(\delta z'_S + \delta z'_{LL}) = \delta z'_Q, $ we can replace $z_S $ with $z'_S = z_S - \delta z'_Q .~$  We find then that the ratio $D_L \, D_S/D_{LS}, $ increases with the intrinsic redshift $ \delta z'_Q, $ from the lens towards the source, which is usually a quasar.

\indent  This is illustrated in columns 6 and 7 of Table 2.  The first column gives the name of the lensing system.
The second and third column give the values of the product in front of brackets in Eq.\,(14) for the big-bang cosmology and in Eq.\,(23) in the plasma-redshift cosmology, respectively.

\indent  From Eq.\,(13), we get for the big-bang cosmology that the value of the product in the second column is
\be
F_{Bb} =(1+z_L) \, \frac{{d_L \, d_S}}{{d_{LS}}} = \frac{{ H_0 }}{{ c }} \, \frac{{ D_L \, D_S }}{{ D_{LS} }}~ ,
\ee
\noindent  where the distances $D$ are obtained from Eq.\,(6) for $(\Omega_k , \, \Omega_m , \, \Omega_{\Lambda} )= (0.0, \, 0.3, \, 0.7),$ and $H_0 = 100.$

\indent  From Eq.\,(21), we get for the plasma-redshift cosmology the value of the product in the third column is
\be
F_{Pl}= \frac{{ H_0 }}{{ c }} \, \frac{{ D_L \, D_S }}{{ D_{LS} }}~ = \frac{{ [{\rm{ln}}(1+z'_L)] \, [{\rm{ln}}(1+z'_S)] }}{{[{\rm{ln}}(1+z'_S)-{\rm{ln}}(1+z'_L)]}}
\ee
\noindent where all the intrinsic redshifts $\delta z' $ are assumed zero, and $H_0 = 100 .$  

\indent The fourth column gives a representative observed value of $H_0 $ as determined in the big-bang cosmology, assuming $(\Omega_k , \, \Omega_m , \, \Omega_{\Lambda} ) = (0.0, \, 0.3, \, 0.7) .$

\indent  The fifth column gives the determination of $H_0$ in the plasma-redshift cosmology assuming no intrinsic redshifts. This value is obtained by multiplying the value of $H_0 ,$ as determined in the big-bang cosmology and listed in the fourth column, by the ratio $[F_{Pl} /F_{Bb}]$ for each system. 

\indent  The sixth column gives the determination of $H_0$ in the plasma-redshift cosmology assuming that the intrinsic redshifts are insignificant except those for the sources, which have intrinsic redshifts of $\delta z'_S + \delta z'_{LL} =  \delta z'_{Q\,1/3} = (z_S - z_L)/3 .$

\indent  The seventh column gives the determination of $H_0$ in the plasma-redshift cosmology assuming that the intrinsic redshifts are insignificant except those for the sources, which have intrinsic redshifts of $\delta z'_S + \delta z'_{LL} =  \delta z'_{Q\,1/2} = (z_S - z_L)/2 .$

\indent  The last row of Table 2 gives the average $H_0$-values.  Column 4 shows that the average value of $H_0 = 62.2$ for $(\Omega_k , \, \Omega_m , \, \Omega_{\Lambda} ) = (0.0, \, 0.3, \, 0.7) .~$  In a big-bang cosmology based on $(\Omega_k , \, \Omega_m , \, \Omega_{\Lambda} ) = (0.0, \, 1.0, \, 0.0) ,$
 which is sometimes used in lensing theory, the $H_0$-values would be 2.8\,\% to 11.1\,\% smaller and the average would be 6.6\,\% smaller, or $H_0 = 58.5 . $

\indent  Column 5 shows that in plasma-redshift cosmology without intrinsic redshifts, the average value $H_0 = 57.9 .~$

\indent  In column 6, we assume that the sources have intrinsic redshifts equal to one third of the redshift difference between the source and the lens.  This corresponds to an intrinsic redshifts in the range of $\delta z_S = 0.092 $ to $ \delta z_S = 1.10 , $ and an average intrinsic redshift of $\delta z_S = 0.463 .~$  The average Hubble constant is then $H_0 = 67.3 .~$  However, there is an indication, as the Table 2 shows, that the source of B0218+357 may have a small or insignificant intrinsic redshift.  This is consistent with observations, which describe this source as a BL Lac object.  The jet from the BL Lac source of the B0218+357 system appears to be pointing at us, and we are then looking at the head of the jet with a much smaller intrinsic redshift than a regular quasar.  The Hubble constant for the B0218+357 system is then $H_0 \approx 67.7 .~$  The average Hubble constant in column 6 is then $H_0 = 65.1 , $ and the average intrinsic redshift of the remaining sources is $\delta z_S = 0.500 .$

\indent  In column 7, we assume that each source has an intrinsic redshift equal to 50\,\% of the redshift difference between the source and the lens.  This corresponds to intrinsic redshifts in the range of $\delta z_S = 0.137 $ to $\delta z_S = 1.65 ,$ and an average intrinsic redshift of $\delta z_S = 0.694 .~$  The average Hubble constant is then $H_0 = 76.6 .~$  However, when we take into account that the source of B0218+357 is a BL Lac object with a nearly insignificant intrinsic redshift, the Hubble constant for the B0218+357 system is $H_0 = 67.7 .~$  The average Hubble constant is then $H_0 = 72.2, $ and the average intrinsic redshift of the remaining sources is $\delta z_S = 0.750. $ 

\indent  The analysis of the data in Table 2 indicates that the intrinsic redshift of quasars is significant.  The intrinsic redshifts could be adjusted to the individual sources to result in a more uniform value of $H_0. $  For example, a system resulting in a low value for the Hubble constant could be assumed to have large intrinsic source redshift, and a system resulting in a high value for the Hubble constant could be assumed to have small intrinsic redshift.  For example, the source of the system B0218+357 appears to have a very small intrinsic redshift or even a small intrinsic blue shift.  This is consistent with the source being a BL Lac object with the light source being a jet pointing at us.


\begin{table}[h]
\centering
{\bf{Table 3.}} \, \, The intrinsic source redshift $\delta z'_Q$ if $(H_0)_{Pl} =70$, and if $(H_0)_{Pl} = 64$
\vspace{2mm}

\begin{tabular}{llllllll}
\hline \hline
&  &  & $(H_0)_{Bb}$ for&$(H_0)_{Pl}$ for& ~~~$\delta z'_Q $ for&~$\delta z'_Q $ for \\
~~~System      & ~~$z_S$&~~$z_L$ &$\delta z'_Q = 0 $ & $\delta z'_Q = 0$&$(H_0)_{Pl}= 70$&$(H_0)_{Pl} = 64 $   \\
\hline \hline
HE1104-1805 &2.319 &0.729 &~~~55.0 &~~~49.4 &~~0.900  &~~0.766  \\
PG1115+080  &1.722 &0.311 &~~~52.5 &~~~50.0 &~~0.978  &~~0.866  \\
SBS1520+530 &1.860 &0.717 &~~~57.0 &~~~51.7 &~~0.547  &~~0.437  \\
B1600+434   &1.590 &0.410 &~~~63.3 &~~~59.7 &~~0.483  &~~0.279  \\
HE2149-2745 &2.030 &0.495 &~~~66.0 &~~~61.1 &~~0.578  &~~0.272  \\
RXJ0911+0551&2.800 &0.770 &~~~59.5 &~~~52.7 &~~1.087  &~~0.873  \\
Q0957+561   &1.410 &0.360 &~~~66.0 &~~~62.9 &~~0.342  &~~0.106  \\
B1608+656   &1.394 &0.630 &~~~69.0 &~~~63.9 &~~0.141  &~~0.016  \\
B0218+357   &0.960 &0.685 &~~~72.3 &~~~67.7 &~~0.018  &$-0.017$  \\
B1422+231   &3.620 &0.330 &~~~64.0 &~~~59.8 &~~2.134  &~~1.464  \\
FBQ0951+2635&1.246 &0.240 &~~~60.0 &~~~58.1 &~~0.511  &~~0.379  \\
\hline
Averages    &      &      &~~~62.2 &~~~57.9 &~~0.702  &~~0.495  \\
\hline \hline
\end{tabular}
\end{table}

\indent  In Table 3, we list in columns 6 and 7 the values of $\delta z'_Q = \delta z'_S + \delta z'_{LL},$ which are obtained by assuming that in the plasma-redshift cosmology $(H_0)_{Pl} \approx 70 $ and $ (H_0)_{Pl} \approx 64 ,$ respectively, for each and every system, and that $ \delta z'_{MW} + \delta z'_{L}\approx 0.003 .~$  

\indent  We show for comparison in the 4th and 5th columns the values of the Hubble constant for the big-bang cosmology and the plasma-redshift cosmology, respectively.  These values are obtained by assuming that $\delta z'_Q \approx 0 .~$  

\indent  The source-redshift and the lens-redshift for each system is listed in the 2nd and the 3rd columns.  The name of each system is shown in the first column.

\indent  From the SNe\,Ia experiments, we found that $ \delta z'_{MW}\approx 0.00095 $ [2], and because all the systems listed in the tables above have Galactic latitude $b > 12^{\circ} ,$ this average value of $ \delta z'_{MW}\approx 0.00095 $ is applicable.  The intrinsic redshift of the lensing galaxy varies with its size and the inclination of its plane to the line of sight.  We assume that the average intrinsic redshift of the lensing galaxy is about twice that for latitudes $b > 12^{\circ} $ at our position in the Milky Way.  We have then that $\delta z'_{L}\approx 0.002 .~$  We get then that $ \delta z'_{MW} + \delta z'_{L}\approx 0.003 .~$  The value $ \delta z'_{MW} + \delta z'_{L}\approx 0.003 $ reduces an average value of $H_0 = 70.5 $ to about $H_0 = 70.0 .~$

\indent  This assumed intrinsic redshift of $\delta z'_{L}\approx 0.002 $ is most likely small for most of the systems.  The values of $\delta z'_{L}$ are likely to be in the range of 0.001 to 0.02 with the higher values applying in case of high density coronas and in case of galaxy clusters.  These higher values for $z'_L$ will lower the $H_0$-values.  For example, had we in the last column of Table 3 used $ \delta z'_{MW} + \delta z'_{L}\approx 0.02 $ instead of 0.003, and used the same values of $\delta z'_Q $ as those listed in column 7 of Table 3, the average value of $ H_0 $ would have been 60.2 instead of 64.  Independent studies comparing the galaxy redshifts with the Tully-Fisher relation [37] have indicated that galaxies may have intrinsic redshifts of about 0.02.

\indent  The intrinsic redshift $\delta z'_{LL} $ will vary from one system to another; but it is most likely larger than  $\delta z'_{L} .~$  Nevertheless, it appears clear that the major portion of the intrinsic redshift of a system, $\delta z'_Q = \delta z'_S + \delta z'_{LL} ,$ is due to the intrinsic redshift, $ \delta z'_S ,$ of the source, which is usually a quasar. 

\indent  Based on more recent estimates for the source redshift of $z_S = 0.944$ [38] for the system B0218+357 and a slightly modified lensing geometry, Wucknitz et al. derived [39] a higher value of $H_0 = 78\pm 6~ {\rm{km\,s}}^{-1} {\rm{Mpc}}^{-1} $ for the Hubble constant.  The corresponding value in plasma-redshift cosmology without any intrinsic redshift is $H_0 = 76.3\pm 6~ {\rm{km\,s}}^{-1} {\rm{Mpc}}^{-1} .~$  The higher estimate for $H_0$ if correct would indicate relatively high average electron density along the line of sight towards the lens, or that the jet from the source is moving towards us causing an intrinsic Doppler blue shift, corresponding to a small negative value in column 7.


\section{Discussion and conclusions}

The main purpose of this analysis is to show in Tables 2 and 3 that the plasma-redshift cosmology is consistent with the present lensing data.  The plasma redshift cosmology predicts and the observations confirm that quasars have large intrinsic redshifts.  Due to their high luminosity, we expect quasars to have hot, dense, and extended coronas, with a large value of $\int N_e \, dx ,$ and therefore relatively large intrinsic redshifts.  Due to the plasma-redshift heating, the corona of the quasar is hot and extends far beyond the Str\"{o}mgren radius of the quasar.  

\indent  The Compton scattering reduces the light intensity from a galaxy with a large redshift of $ z = 3 $ by a factor of $ 1/[1+z]^2  = 1/16 .~$  Similarly, the Compton scattering reduces the light intensity from a quasar with a redshift of $z = 3$, including an intrinsic redshift of $ \delta z'_S = 2 $, by a factor of only $ 1/[1+(z-2)]^2 = 1/4 ,$ because most of the light scattered in the corona of the quasar would be observed as coming from the quasar.  The ratio $(D_Q/D_G)^2 ,$ where $D_Q$ and $D_G$ are the distances given by Eq.\,(4), would result in another factor of 1/4.  If the two objects actually emitted the same photon flux, the quasar would appear about 16 times (3 magnitudes) brighter than the galaxy.  A quasar with a redshift of 6.4 and intrinsic redshift of 5.4 or 5.9 would similarly appear in the big-bang cosmology to be 114 and 593 times brighter than its actual absolute brightness.  The extended, hot, and dense corona can help explain simultaneously the x-ray, infrared, and microwave luminosity of these objects.  The high-energy gamma rays (often up to several TeV and leading to particle creation) must also be considered [1].  The intrinsic redshift caused by this corona explains also the rapid change in luminosity of many quasars free of time dilation.  Plasma redshift leads thus to a reasonable explanation of many phenomena that have been difficult to explain.

\indent  Future improvements should make it possible to obtain better estimates of the intrinsic redshifts from observing the small shifts and widths of the absorption lines.  For example, when the light from the source passes through the corona around the lensing galaxy, we may be able to observe the widths and small shifts of the different lines to help us estimate the densities, temperature and the extent of the corona.

\indent  It is of interest to note that the system B0218+357, which we knew should have a small intrinsic shift, is the only system found by this analysis to have a small intrinsic source redshift and even a small blue shift.

\indent  We consider this analysis as demonstrating 1) that the plasma redshift is consistent with the observations, and 2) that the quasars really have large intrinsic redshifts.

\indent  Great many observations have indicated intrinsic redshifts of bright stars, galaxies, and quasars [27].  However, we have never had a reliable theory for explaining these observations.  The plasma redshift theory is reliable, as it is based on well-established physics [1].  It predicts that quasars and galaxies must have significant intrinsic redshifts.


\end{document}